\documentclass[twocolumn,showpacs,prl,amsmath,amssymb,superscriptaddress]{revtex4}
\usepackage{color}
\usepackage{float}
\usepackage{verbatim}
\usepackage{calc}
\usepackage{float}
\usepackage{subfigure} 
\usepackage{graphicx}
\usepackage{dcolumn}
\usepackage{bm}

\begin{document}

\title{Structural Transitions in Dense Networks}

\author{R. Lambiotte}
\affiliation{naXys, Namur Center for Complex Systems, University of Namur,
rempart de la Vierge 8, B 5000 Namur, Belgium}
\author{P. L. Krapivsky}
\affiliation{Department of Physics, Boston University, Boston, Massachusetts 02215, USA}
\author{U. Bhat}
\affiliation{Department of Physics, Boston University, Boston, Massachusetts 02215, USA}
\affiliation{Santa Fe Institute, 1399 Hyde Park Road, Santa Fe, NM, 87501}
\author{S. Redner}
\affiliation{Santa Fe Institute, 1399 Hyde Park Road, Santa Fe, NM, 87501}

\begin{abstract}

  We introduce an evolving network model in which a new node attaches to a
  randomly selected target node and also to each of its neighbors with
  probability $p$.  The resulting network is sparse for $p<\frac{1}{2}$ and
  dense (average degree increasing with number of nodes $N$) for
  $p\geq \frac{1}{2}$.  In the dense regime, individual networks realizations
  built by this copying mechanism are disparate and not self-averaging.
  Further, there is an infinite sequence of structural anomalies at
  $p=\frac{2}{3}$, $\frac{3}{4}$, $\frac{4}{5}$, etc., where the dependences
  on $N$ of the number of triangles (3-cliques), 4-cliques, undergo phase
  transitions.  When linking to second neighbors of the target can occur, the
  probability that the resulting graph is complete---where all nodes are
  connected---is non-zero as $N\to\infty$.

\end {abstract}

\maketitle

The investigation of complex networks has blossomed into a rich discipline,
with many theoretical advances and a myriad of applications to the physical
and social sciences~\cite{AB02,DM03,N03,NBW06,BBV08,N10}.  Much of the focus
has been on \emph{sparse} networks, where the average degree, defined as the
average number of links attached to a node, is finite as the number of nodes
in the network $N\to\infty$.  In this letter, we introduce a minimal
generative model for \emph{dense} networks, in which the average degree grows
with $N$.

In addition to the many new phenomena that arise in the dense regime, such
networks may account for structural properties of the brain, which, for
humans, has average degree $10^3$ ($10^{11}$ neurons, $10^{14}$
interconnections).  The brain exhibits a rich spectrum of motifs---small
subsets of densely interconnected nodes~\cite{HK04,ECCBA05,NLFEB09} that may
underlie its wondrous functionality.  Densification also appears to arise in
many empirical networks, including, for example, the arXiv citation, patent
citation, and autonomous systems graphs~\cite{kleinberg}.  As we present
below, our model displays many intriguing features that may mirror some of
these structural properties, including a sequence of phase transitions in the
densities of fixed-size cliques (complete subgraphs), non-extensivity of the
degree distribution, and lack of self averaging.

\begin{figure}[ht]
\begin{center}
\includegraphics[width=0.4\textwidth]{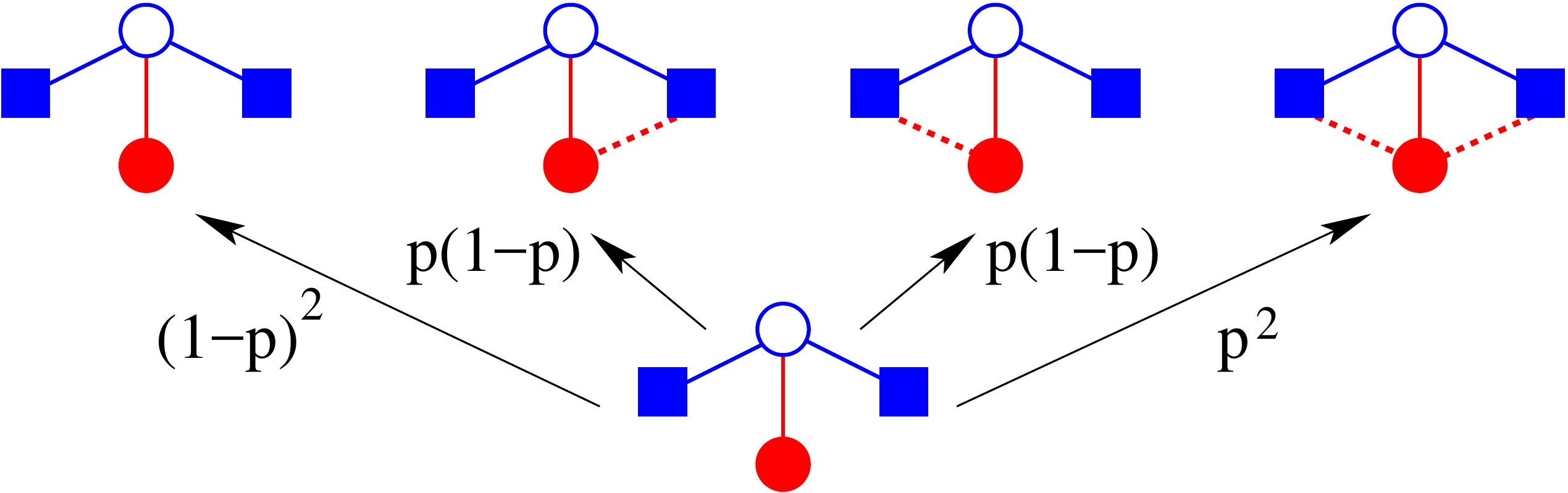}
\caption{\small The copying mechanism.  A new node (filled circle) attaches
  to a random target (open circle) and to each of the friends of the target
  (squares) with probability $p$. }
\label{cartoon}
\end{center}
\end{figure}

Our model is based on the generic mechanism of copying (see also
Ref.~\cite{kleinberg}): new nodes are introduced sequentially and each
connects to a random pre-existing target node, as well as to each the
neighbors of the target (friends of a friend) independently with probability
$p$ (Fig.~\ref{cartoon}).  This mechanism drives the dynamics of social
networks~\cite{granovetter,toivonen,AA16}, as well as social media, such as
Facebook, where people are invited to connect to a friend of a friend (see,
e.g.,~\cite{JR07,KR05}).  Copying is also related to triadic
closure~\cite{R53,HK02,DES02,TOSHK06,BDIF14,BKR14}, which naturally generates
highly clustered networks.

The copying and related redirection mechanisms are ubiquitous in networks;
they underlie the world-wide web, citation, and other information
networks~\cite{kleinberg,GNC,sergi,evans}, the evolutionary process of gene
duplication~\cite{japan,bio}, and protein interaction
networks~\cite{pastor,chung,vesp,korea,wag,nik,protein,sym}.  Finally,
copying is local~\cite{vaz,GNR,saram}, as the creation of new links only
depends on the local neighborhood of the target node, contrary to global
rules such as preferential attachment~\cite{AB02,DM03,N03,NBW06,BBV08,N10}.
As we will show, copying leads to highly non-trivial networks, but the
simplicity of this mechanism allows for analytical solution for many network
properties.

\begin{figure*}
\hskip 0.4in\includegraphics[width=0.15\textwidth]{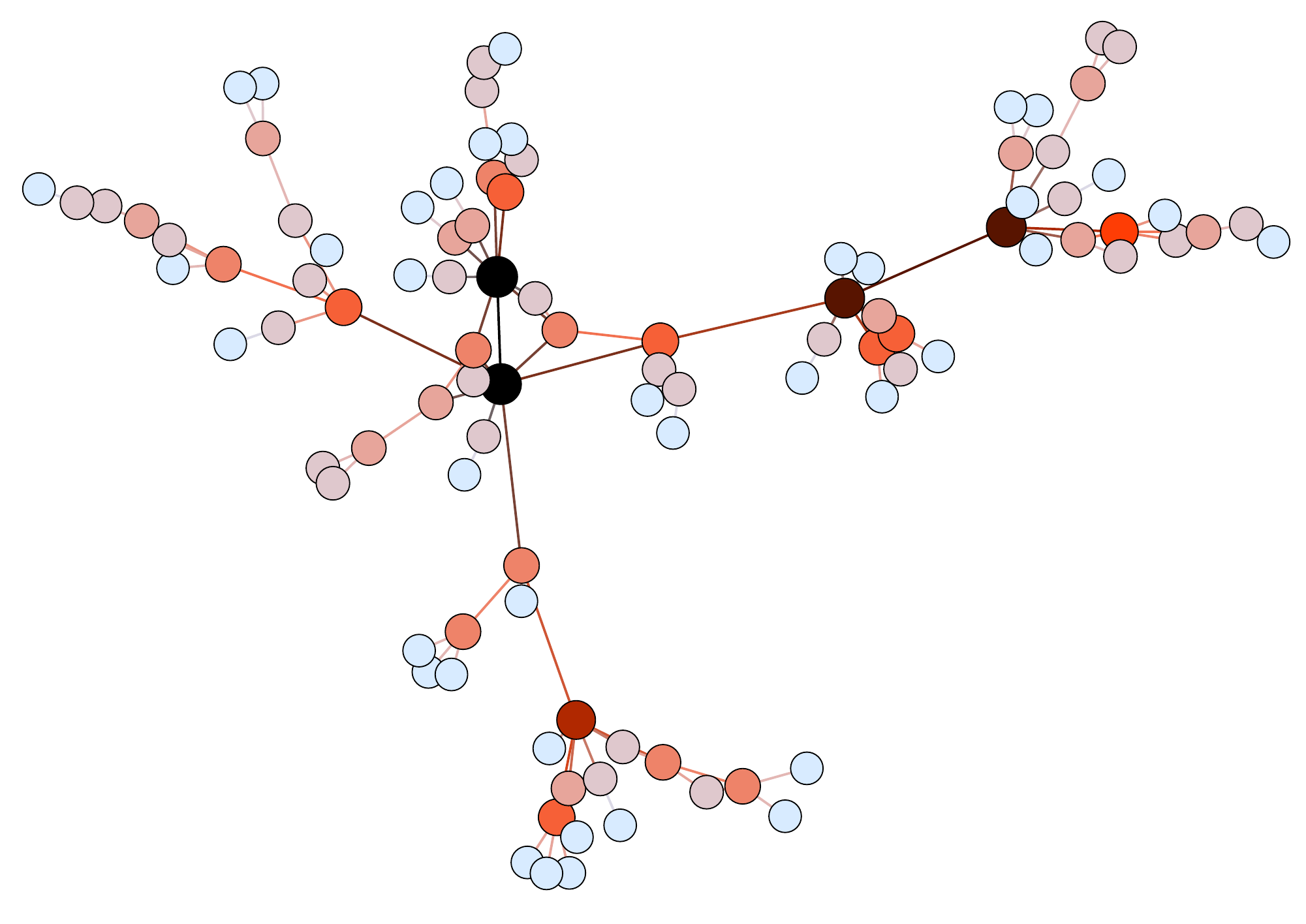}
\hskip 0.15in\includegraphics[width=0.15\textwidth]{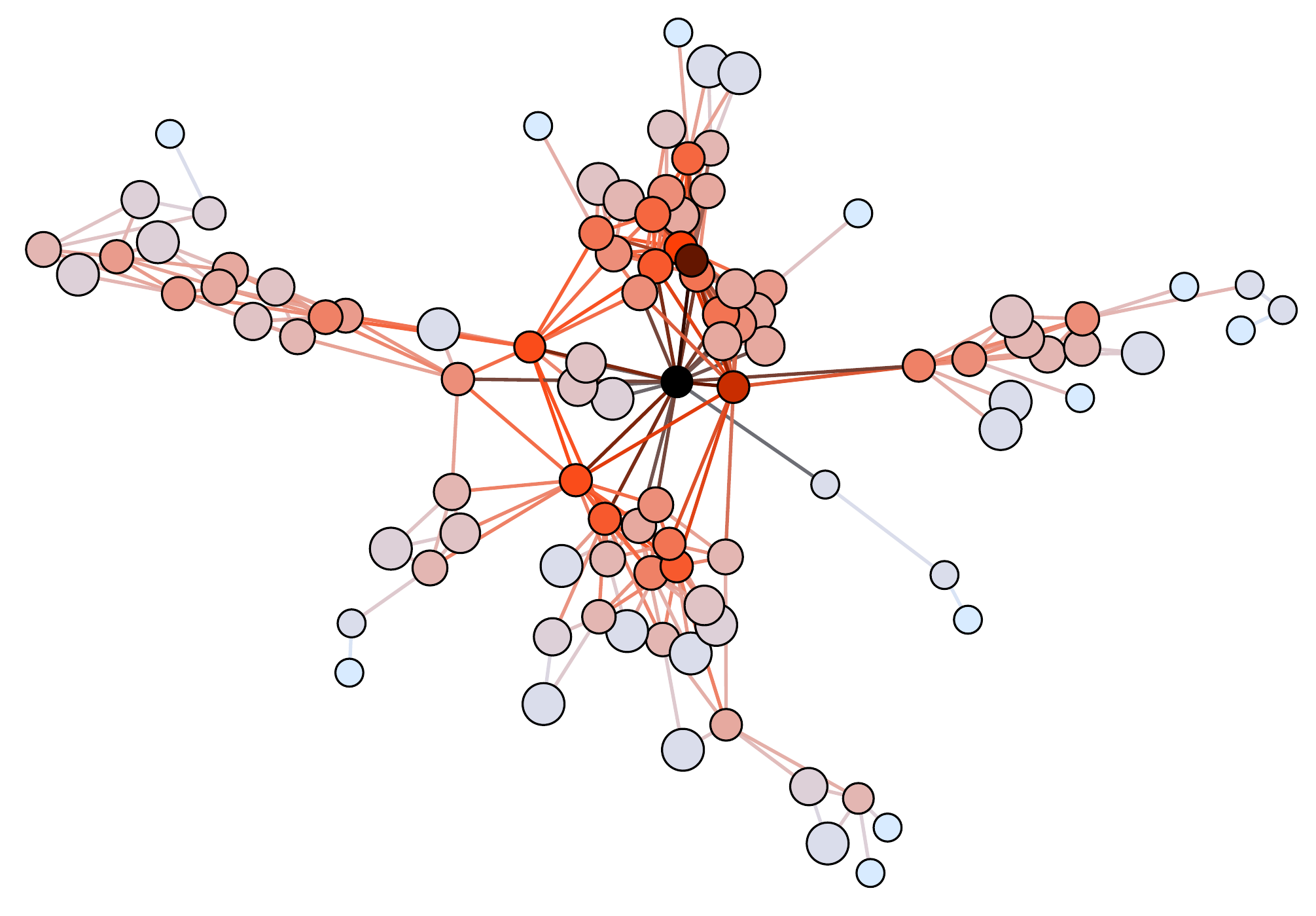}
\hskip 0.3in\includegraphics[width=0.12\textwidth]{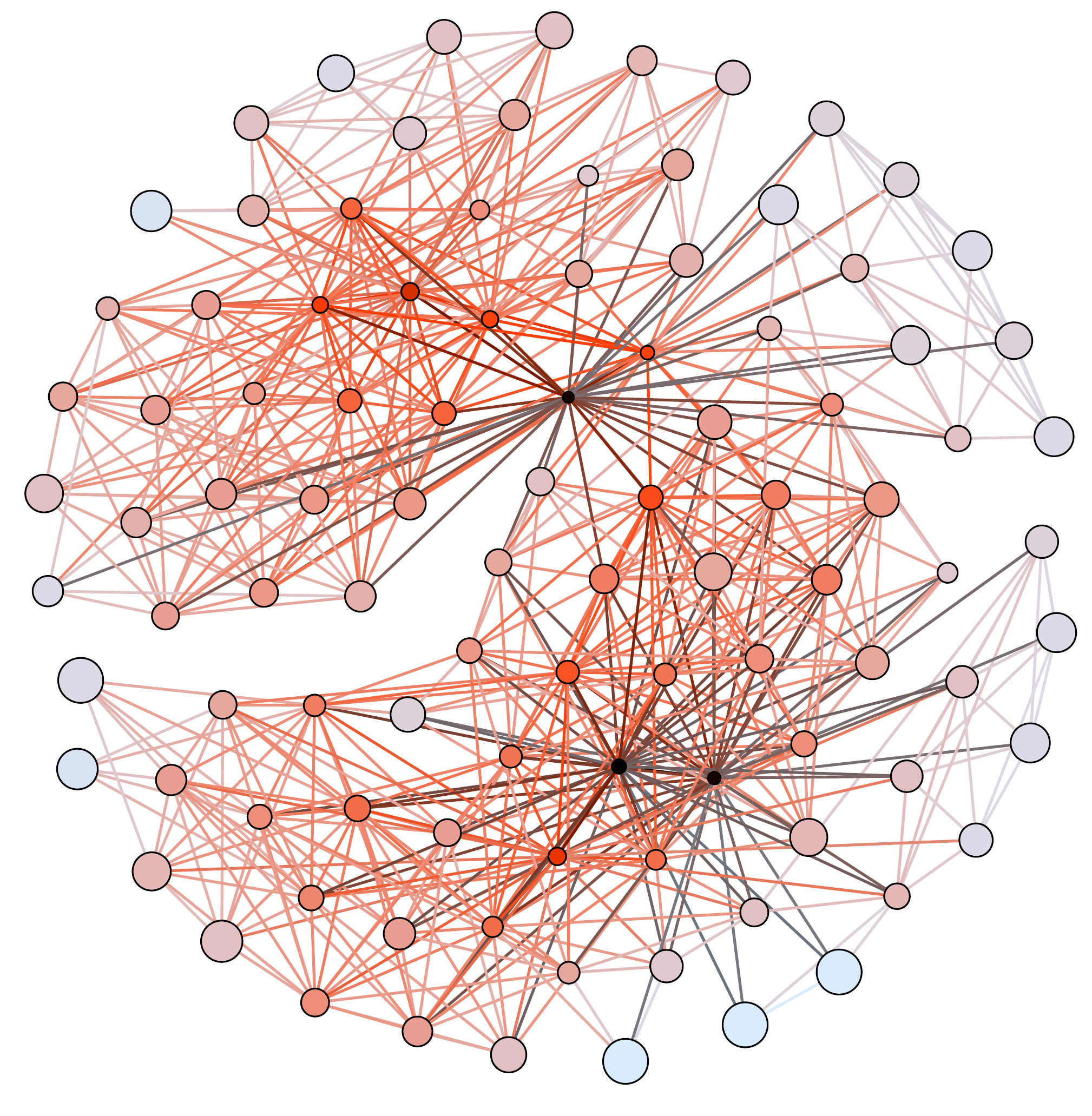}
\hskip 0.4in\includegraphics[width=0.12\textwidth]{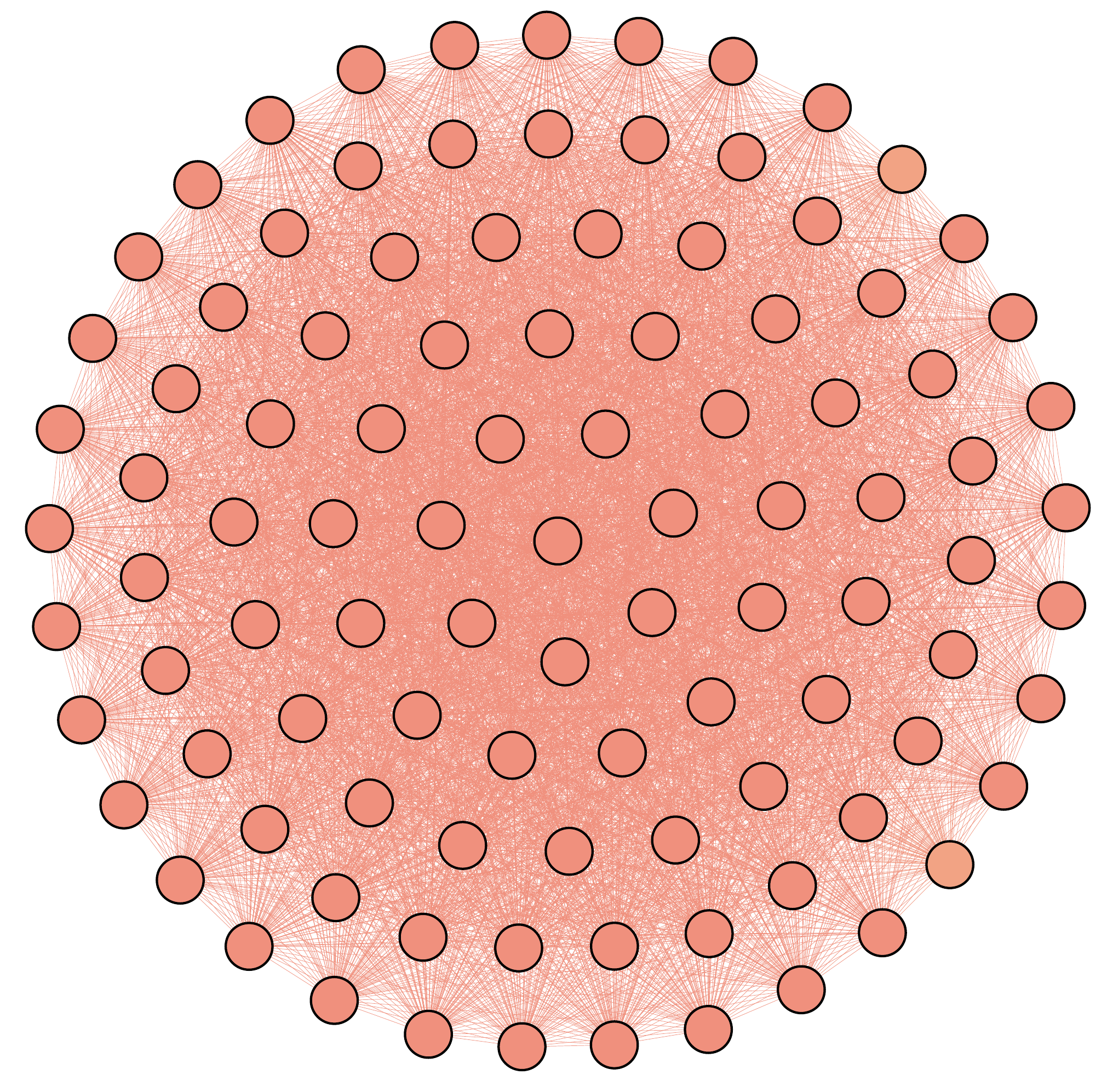}
\hbox{\hskip 1.1in\includegraphics[width=0.8\textwidth]{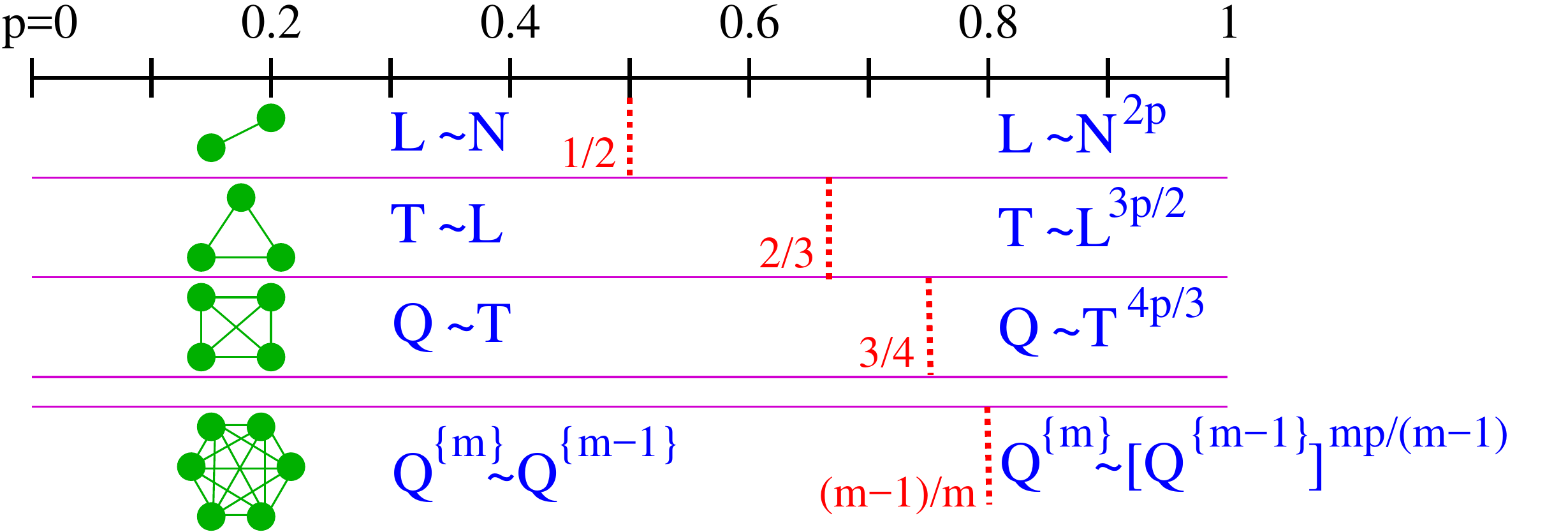}}
\caption{Realizations of the copying model for $p=0.1$, $0.4$, 0.7, and 1 for
  $N=100$, and a summary of the dense regimes.}
\label{vis}
\end{figure*}

When $p = 0$, a network built by copying is a random recursive
tree~\cite{rrt1,rrt2,rrt3}, while for $p = 1$, a complete graph arises if the
initial graph is also complete.  For $p<\frac{1}{2}$, the network is sparse,
while for $p\geq \frac{1}{2}$, the number of links grows superlinearly with
$N$ and the network is \emph{dense}.  In the dense regime the network is
highly clustered (Fig.~\ref{vis}) and undergoes an infinite series of
structural transitions at $p=\frac{2}{3}, \frac{3}{4}, \frac{4}{5},\ldots$
that signal sudden changes in the growth laws of the number of 3-cliques
(triangles), 4-cliques (tetrahedra), etc.
  
\smallskip
\noindent{\tt Number of Links.}
\label{total}
We first investigate how copying affects the growth in the number of
links.  Let $L_N$ denote the average number of links in a network of $N$
nodes.  Adding a new node increases the number of links, on average, by
$1+p\langle k\rangle$, where $\langle k\rangle=2L_N/N$ is the average degree.
Thus $L_N$ grows according to
\begin{equation}
\label{LN-exact}
L_{N+1}=L_N+1+2p\,\frac{L_N}{N}.
\end{equation}
Taking the continuum $N\gg 1$ limit and solving the resulting differential
equation gives
\begin{equation}
\label{LN-sol}
L_N= 
\begin{cases}
N/(1-2p)      & \qquad p<\tfrac{1}{2},\cr
N\,\ln N          &  \qquad p=\tfrac{1}{2},\cr
A(p)\,N^{2p}      & \qquad \tfrac{1}{2}<p\leq 1,
\end{cases} 
\end{equation}
with amplitude $A(p)=\big[(2p-1)\Gamma(1+2p)\big]^{-1}$ that is calculable by
solving the discrete recursion \eqref{LN-exact}.  Indeed, the recurrence
\eqref{LN-exact} admits the exact solution
\begin{equation*}
L_N=\frac{\Gamma(2p+N)}{\Gamma(N)}\sum_{j=2}^{N}\frac{\Gamma(j)}{\Gamma(2p+j)}.
\end{equation*}
from which the asymptotics \eqref{LN-sol} and the amplitude $A(p)$ follow~\cite{LKBR16}.

We can also compute~\cite{LKBR16} the standard deviation
$\Sigma_L\equiv\sqrt{\langle L_N^2\rangle-\langle L_N\rangle^2}$, which
exhibits an even richer dependence on $N$, with transitions at
$p=\frac{1}{4}$ and $p=\frac{1}{2}$:
\begin{equation}
\label{VN-sol}
\Sigma_L \sim  
\begin{cases}
\sqrt{N}               &  \qquad p<\tfrac{1}{4},\cr
\sqrt{N\,\ln N}      & \qquad p=\tfrac{1}{4},\cr
N^{2p}       & \qquad \tfrac{1}{4}<p<1, p\ne \tfrac{1}{2},\cr
N\,\sqrt{\ln N}   &\qquad p=\tfrac{1}{2}. 
\end{cases} 
\end{equation}
The salient consequences of \eqref{LN-sol} and \eqref{VN-sol} are that $L_N$
grows superlinearly with $N$ and is {\em not} self averaging for
$p>\frac{1}{2}$.  This feature implies that there is a wide diversity among
different network realizations, and the first few steps are crucial in
shaping the evolution.  Conversely, fluctuations are negligible in the sparse
phase, where $\Sigma_L/\langle L_N\rangle\to 0$ as $N\to\infty$.  Only for
$p<\tfrac{1}{4}$, where $\Sigma_L$ scales as $\sqrt{N}$, do we anticipate
that the distribution $P(L_N)$ is asymptotically Gaussian.

\begin{figure}[h]
\begin{center}
\includegraphics[width=0.2\textwidth]{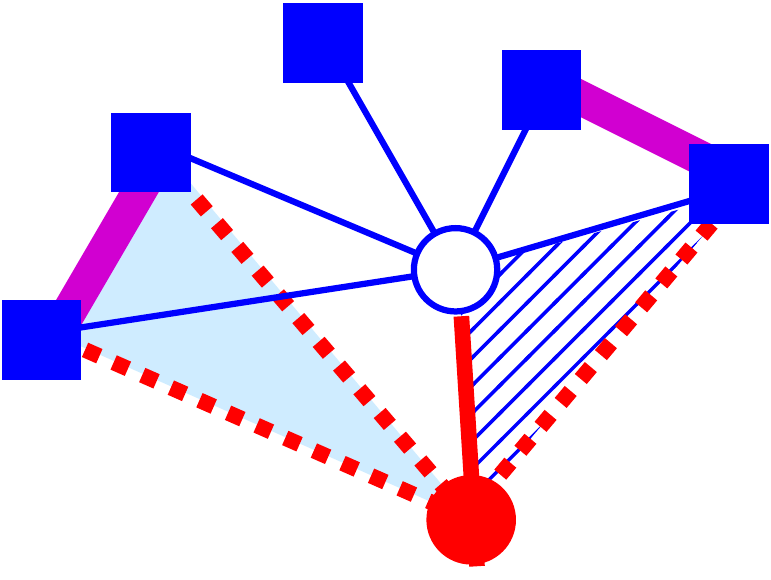}
\caption{\small Counting triangles.  The target node (open circle) has five
  neighbors (squares), two of which are joined by `clustering' links (heavy
  lines).  Three copying links (dashed) create three new triangles (one is
  hatched for illustration) and one new triangle by induced linking
  (shaded).}
\label{TN}
\end{center}
\end{figure}

\smallskip
\noindent{\tt Triangles and Larger Cliques.}
A related set of transitions occurs in the densities of larger-size cliques.
A $k$-clique is a complete subgraph of of $k(k-1)/2$ links that completely
connect $k$ nodes within the network.  We first investigate the number of
3-cliques (triangles).  There are two mechanisms that increase the number of
triangles as a result of a copying event---direct and induced linking.  In a
direct linking, a triangle is created in each copying event that consists of
the new node, the target node, and the neighbor of the target that receives a
`copying' link (Fig.~\ref{TN}).  In an induced linking, additional triangles
are created whenever copying creates links to more than one neighbor of the
target that were previously linked.

To determine the $N$-dependence of average number of triangles, suppose that
the target node has degree $k$ and that its neighbors are connected via $c$
`clustering' links (Fig.~\ref{TN}).  If $a$ copying links are made, the
number of triangles increases on average by
\begin{equation}
\label{deltaT}
\Delta T = a + \frac{a (a-1)}{2} \frac{c}{k (k-1)/2}.
\end{equation}
The first term on the right accounts for direct linking and the second for
induced linking.  For the latter, we need to count how many of $a(a-1)/2$
possible links between $a$ neighbors of the target, which also connect to the
new node, are actually present.  Averaging \eqref{deltaT} with respect to the
binomial distribution for $a$, we obtain, after an elementary calculation,
\begin{equation}
\label{deltaT2}
\overline{\Delta T} =  pk + p^2 c.
\end{equation}
The term $p^2 c$ arises because two connected neighbors of the target also
connect to the new node with probability $p^2$, since linking to each node
occurs independently.

We now express the average number of clustering links $\langle c \rangle$ in
terms of the number of triangles $T_N$.  To this end, we note that $c$ equals
the number of triangles that contain the target node,
$\langle c \rangle = 3 T_N/N$.  Using these relations, the average number of
triangles increases by
$\langle \overline{\Delta T_N} \rangle =3 p^2 T_N/N + 2 p L_N/N $ with each
node addition.  For $N \gg 1$, we thus obtain the rate equation
\begin{equation}
\label{ctevolution}
\frac{dT_N}{dN} = 3 p^2 \frac{T_N}{N}  + 2p \frac{L_N}{N}\,,
\end{equation}
whose solution is
\begin{equation}
\label{TN-sol}
T_N= 
\begin{cases}
\frac{2p}{(1-2p)\,(1-3p^2)}\, N      & \qquad p<\tfrac{1}{2},\cr
4 N\,\ln N                                    & \qquad p=\tfrac{1}{2},\cr
\frac{A(p)}{1-3p/2} \,N^{2p}      & \qquad \tfrac{1}{2}<p < \tfrac{2}{3},\cr
\frac{4}{\Gamma(4/3)} \,N^{4/3} \ln N     & \qquad p=\tfrac{2}{3},\cr
B(p)\,N^{3p^2}      & \qquad \tfrac{2}{3} <p\leq 1,
\end{cases} 
\end{equation}
with $A(p)$ given in \eqref{LN-sol} and $B(p)$ also calculable~\cite{LKBR16}
by solving the discrete recursion for $T_N$.

Equation~\eqref{TN-sol} shows that the average number of triangles $T_N$
undergoes a second phase transition at $p=\frac{2}{3}$ where $T_N$ grows
superlinearly in $L_N$ (Fig.~\ref{vis}).  Moreover, the density of triangles
converges to a non-vanishing value when $0 < p<\tfrac{1}{2}$, which mirrors
the high density of triangles found in many complex
networks~\cite{R53,HK02,DES02,TOSHK06,BDIF14,BKR14}.

The reasoning presented above can be generalized to 4-cliques (quartets) and
we find that their number grows according to the rate equation~\cite{LKBR16}
\begin{equation}
\label{cqevolution}
\frac{dQ_N}{dN} = 3 p^2 \frac{T_N}{N}  + 4 p^3 \frac{Q_N}{N},
\end{equation}
from which, the average number of quartets grows as (with all prefactors
omitted)
\begin{equation*}
\label{QN-sol}
Q_N \sim
\begin{cases}
  N & \qquad \hskip 0.05cm 0< p<\tfrac{1}{2},\\
  N^{2p} & \qquad \tfrac{1}{2}<p < \tfrac{2}{3},\\ 
  N^{3 p^2} & \qquad   \tfrac{2}{3}<p < \tfrac{3}{4},\\
   N^{4 p^3} & \qquad \tfrac{3}{4}<p\leq 1.
\end{cases} 
\end{equation*}
At the transition points $p = \frac{1}{2}$, $\frac{2}{3}$, and $\frac{3}{4}$,
the algebraic factor is multiplied by $\ln N$.

Generally, the average number $Q^{\{m\}}_N$ of $m$-cliques evolves according to
\begin{equation}
\label{cmevolution}
\frac{dQ^{\{m\}}_N}{dN} = (m-1) p^{m-2} \frac{Q^{\{m-1\}}_N}{N}  + m p^{m-1} \frac{Q^{\{m\}}_N}{N}\,.
\end{equation}
This behavior is analogous to what occurs in duplication-divergence
networks~\cite{sym}.  Solving \eqref{cmevolution} recursively gives
\begin{equation}
\label{Mm}
Q^{\{m\}}_N \sim N^{(j+1) p^j}      \qquad   \tfrac{j}{j+1}<p < \tfrac{j+1}{j+2}\,,
\end{equation}
with $j = 0, 1, 2,\ldots, m-1$.  Thus the $N$-dependence of the average number
of cliques of size $m$ undergoes transitions at $p = 1 - \frac{1}{n}$ with
$n = 2, \ldots , m$.

\smallskip
\noindent{\tt Degree Distribution.}  Let $N_k$ be the number of nodes of
degree $k$.  Following standard reasoning~\cite{GNR,korea}, the degree
distribution evolves according to
\begin{subequations}
\label{Nk-both}
\begin{equation}
\label{NkN}
\frac{dN_k}{dN} = \frac{N_{k-1}\!-\!N_k}{N}+p\,\frac{(k\!-\!1)N_{k-1}\!-\!kN_k}{N}+m_k.
\end{equation}
The first term on the right is the contribution due to attachment to the
target node, the second term accounts for attachments to the neighbors of the
target node, and the third term
\begin{equation}
\label{mk}
m_k\equiv\sum_{s\geq k-1} n_s
\binom{s}{k-1} p^{k-1} (1-p)^{s-k+1}
\end{equation}
\end{subequations}
is the probability that the new node acquires a degree $k$ because it
attaches to a target of degree $s$ and to $k-1$ neighbors of this target.
Here $n_s=N_s/N$ denotes the fraction of nodes of degree $s$.

When the network is sparse and large, we assume that the fractions $n_k$ do
not depend on $N$ to recast \eqref{Nk-both} to
\begin{equation}
\label{nk-long}
[2+p(k\!+\!1)]n_{k+1}\!=[1+pk]n_k
+\!\sum_{s\geq k}\! n_s \binom{s}{k} p^{k} (1\!-\!p)^{s-k}.
\end{equation}
This equation is not a recurrence, but it is still possible to extract its
asymptotics.  First, we observe that for large $k$, the summand on the right
is sharply peaked around $s\approx k/p$ and thus reduces
to~\cite{korea,protein}
\begin{equation*}
n_{k/p}\sum_{s\geq k} 
\binom{s}{k} p^{k} (1-p)^{s-k}=p^{-1}n_{k/p}\,,
\end{equation*}
where we used a binomial identity to compute the sum itself.  Substituting
this into Eq.~\eqref{nk-long} and assuming that $n_k$ decays slower than
exponentially so that differences may be replaced by derivatives, we obtain
the non-local equation for the degree distribution
\begin{equation}
\label{nk-eq}
\frac{d}{dk}\,[1+pk]n_k=p^{-1}\,n_{k/p} - n_k\,.
\end{equation}
The algebraic form $n_k\sim k^{-\gamma}$ solves this equation and also gives
the transcendental relation for the exponent,
\begin{equation}
\label{gamma}
\gamma=1+p^{-1}-p^{\gamma-2}\,,
\end{equation}
which admits two solutions.  One, $\gamma=1$, is unphysical because the
corresponding degree distribution is not normalizable.  The other applies
when $0\leq p<\frac{1}{2}$, where $\gamma=\gamma(p)$ decreases monotonically
with $p$, with $\gamma(0)=\infty$ and $\gamma(\tfrac{1}{2})=2$.  Because
$\gamma>2$ for $0\leq p<\frac{1}{2}$ the average degree
$\langle k\rangle=\sum_{k\geq 1}kn_k$ is finite so that the network is indeed
sparse for $0\leq p<\frac{1}{2}$.

\begin{figure}
\includegraphics[width=0.4\textwidth]{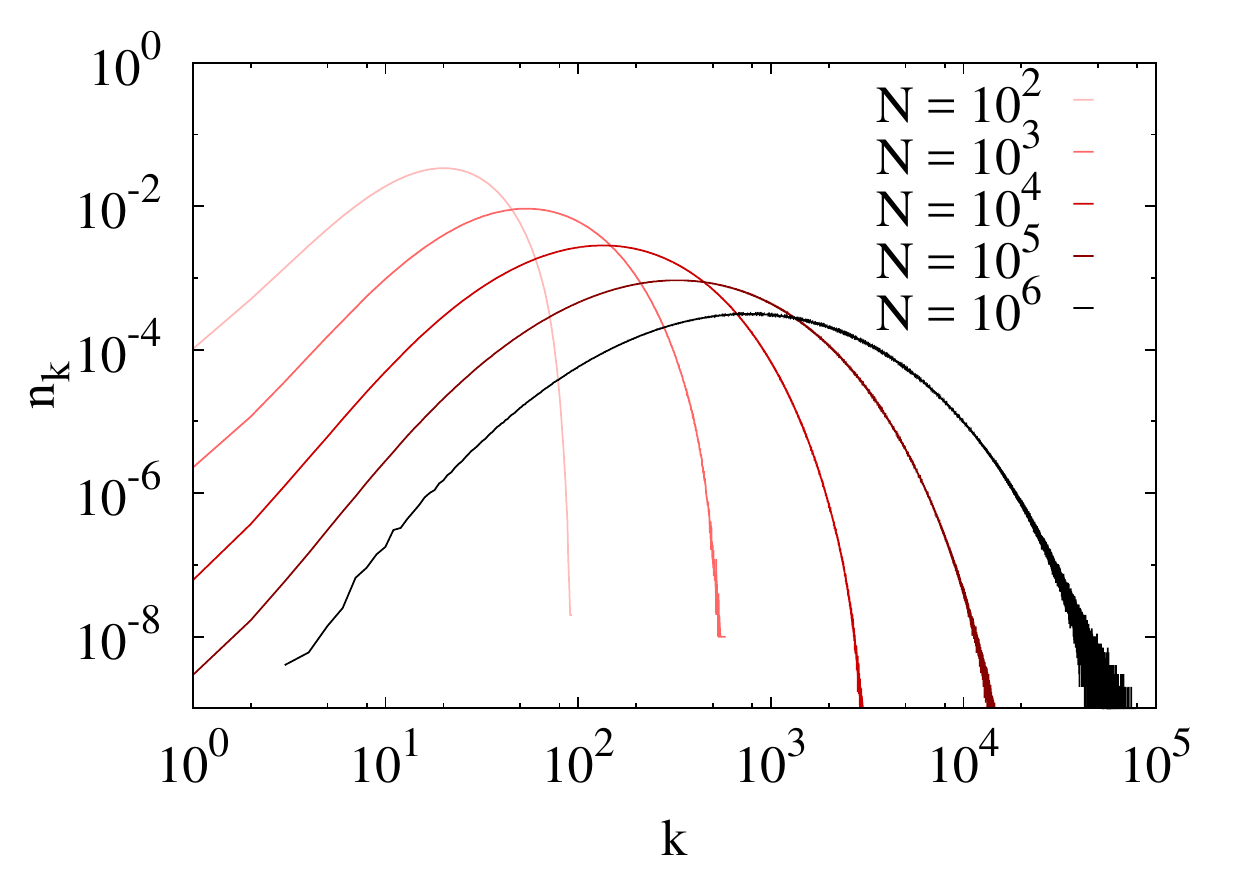}
\caption{Simulations of $10^{10}/N$ realizations for the degree distributions
  $n_k$ for $p=0.75$ (dense regime) and various $N$.}
\label{deg-dist}
\end{figure}

In the dense regime, the analysis above no longer applies and we resort to
simulations.  We find that the degree distribution is no longer stationary;
that is, $n_k$ depends on $N$, in contrast to the stationarity in the sparse
regime.  Moreover, the degree distribution appears to slowly converge to a
form that is close to log-normal as $N\to\infty$ (Fig.~\ref{deg-dist}).

\smallskip
\noindent{\tt Network Completeness.} Suppose that in addition to connecting
to the neighbors of the target with probability $p$, a new node \emph{also}
connects to the second neighbors of the target with probability $p_2$.  Such
a mechanism naturally arises in social media, where connections to friends of
a friend can extend to higher-order acquaintances.  The surprising feature of
second-order linking is that the network is complete with non-zero
probability.

Let $\Pi_N$ denote the probability that a network of $N$ nodes always remains
complete for connection probabilities $p$ and $p_2$.  This completeness
probability is
\begin{equation}
\label{PN}
  \Pi_N = \prod_{r=1}^{N-1} \sum_{k=0}^{r-1} \mathrm{B}(r, k, p)
  \left(1-(1-p_2)^{k}\right)^{r-k-1}\,,
\end{equation}
where $\mathrm{B}(r, k, p)=\binom{r-1}{k} p^{k}(1-p)^{r-k-1}$ is the binomial
probability that copying leads to $k$ links to the neighbors of the target.
The second factor is the probability that all of the remaining $r-k-1$
neighbors of the target are connected by second-order links.

\begin{figure}
\includegraphics[width=0.4\textwidth]{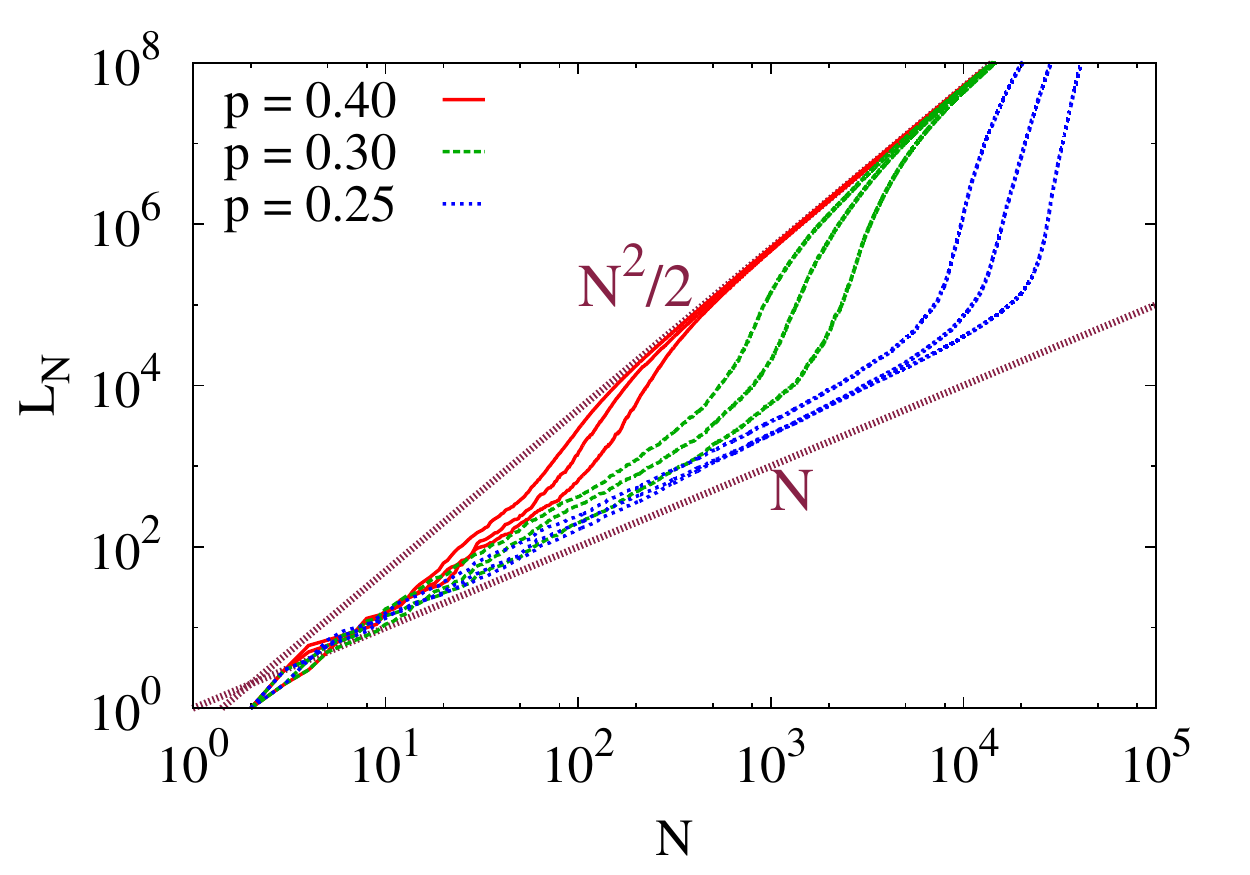}
\caption{The $N$ dependence of the number of links for second-neighbor
  copying with $p_2=p^2$.}
\label{L-vs-N}
\end{figure}

Asymptotic analysis and numerical evaluation of \eqref{PN} show that $\Pi_N$
indeed converges to a non-zero, albeit extremely small, value~\cite{LKBR16}.
A more relevant criterion is not defect-free completeness, but whether the
number of links eventually scales as $N^2/2$, as in the complete graph.
Simulations show that for reasonable values of $p$ and $p_2$, $L_N$ initially
grows linearly with $N$ but then crosses over to growing as $N^2/2$
(Fig.~\ref{L-vs-N}).  Thus second-order copying generically leads to networks
that are effectively complete---eventually each individual knows almost
everybody.  Moreover Fig.~\ref{L-vs-N} illustrates the macroscopic
differences between individual network realizations.  Thus copying leads to
non-self-averaging in the dense regime---unpredictable outcomes when starting
from a fixed initial state.  This intriguing feature also arises in empirical
networks and related models~\cite{SDW06,RW12,O14}, and intellectually
originates with the classic P\'olya urn model~\cite{M17,EP23,M09}.

To summarize, we introduced a simple and rich generative model for dense
networks based on the copying mechanism.  The resulting network is dense for
copying probability $p\geq \tfrac{1}{2}$.  The dense regime partitions into
distinct windows where the density of $k$-cliques each have unique scaling
properties.  Generally, different realizations of the network in the dense
regime are extremely diverse.  The degree distribution appears to slowly
converge to a log-normal form in the dense regime.  When second-neighbor
connections can occur, the network asymptotically becomes complete.

Financial support for this research was also provided in part by the grants
from the ARC and the Belgian Network DYSCO, funded by the IAP Programme (RL),
DMR-1608211 and 1623243 from the National Science Foundation (UB and SR), the
John Templeton Foundation (SR), and Grant No.\ 2012145 from the United States
Israel Binational Science Foundation (UB).

\end{document}